\documentclass[fleqn,twoside,twocolumn,nofootinbib,showkeys]{revtex4} 
\usepackage[nocpr]{ujp} 

\begin{document}
\title[In-Plane Anisotropy Effect on Critical Transition Field]
{IN-PLANE ANISOTROPY EFFECT ON CRITICAL\\ TRANSITION FIELD IN
NANOGRANULAR FILMS\\
WITH PERPENDICULAR ANISOTROPY}
\author{M.M.~Kulyk}
\affiliation{Institute of Physics, Nat. Acad. of Sci. of Ukraine}
\address{46, Nauky Ave., Kyiv 03680, Ukraine}
\email{nikolaj.kulik.ifnasu@gmail.com}
\author{V.M.~Kalita}%
\affiliation{Institute of Physics, Nat. Acad. of Sci. of Ukraine}%
\address{46, Nauky Ave., Kyiv 03680, Ukraine}%
\email{nikolaj.kulik.ifnasu@gmail.com}
\author{A.F.~Lozenko}
\affiliation{Institute of Physics, Nat. Acad. of Sci. of Ukraine}
\address{46, Nauky Ave., Kyiv 03680, Ukraine}
\author{S.M.~Ryabchenko}%
\affiliation{Institute of Physics, Nat. Acad. of Sci. of Ukraine}%
\address{46, Nauky Ave., Kyiv 03680, Ukraine}%
\email{nikolaj.kulik.ifnasu@gmail.com}
\author{O.V.~Stognei}
\affiliation{Voronezh State Technical University}
\address{14, Moscow Ave., Voronezh 394026, Russian Federation}
\author{A.V.~Sitnikov\,}%
\affiliation{Voronezh State Technical University}%
\address{14, Moscow Ave., Voronezh 394026, Russian Federation}%
 \udk{537.624.9, 537.622.4,\\[-3pt] 537.9} \pacs{75.70.Ak, 75.30.Gw,\\[-3pt] 75.60.Ej, 62.23.Pq} \razd{}

\autorcol{M.M.\hspace*{0.7mm}Kulyk, V.M.\hspace*{0.7mm}Kalita,
A.F.\hspace*{0.7mm}Lozenko et al.}

\setcounter{page}{52}%

\begin{abstract}
The influence of the in-plane anisotropy on the magnetization of a
nanogranular film with perpendicular anisotropy has been
studied.\,\,It is shown that if a magnetic field is tilted with
respect to the film normal, a critical transition from the
inhomogeneous magnetic state of granules with noncollinear
directions of their moments to the homogeneous one with parallel
orientation of granular magnetic moments takes place.\,\,The
in-plane anisotropy is found to affect the angular dependence of the
critical field.\,\,The ensemble of oriented biaxial particles is
theoretically described in the double-well potential approximation.
Despite the biaxial magnetic anisotropy of particles, their
ensemble, if in the inhomogeneous state, is divided into two
subensembles, with the magnetic moments of particles being collinear
in each of them.\,\,In the critical field, a transition from the
inhomogeneous state with two subensembles into the homogeneous one
takes place.\,\,The results of theoretical calculations are compared
with experimental data for a nanogranular Co/Al$_{2}$O$_{n}$ film
with perpendicular anisotropy containing 74.5~at.\%~Co, which
exceeds the percolation threshold.\,\,The magnetic moment of this
film is a sum of two contributions: from nanogranules with biaxial
anisotropy and a phase forming the percolation cluster.\,\,The
magnetic properties of nanogranules, whose contribution is separated
from the total film magnetization, agree well with the calculation
data.
\end{abstract}


\maketitle

\section{Introduction}

Nanogranular magnetic films attract a keen interest owing to the
prospects of their application in various functional devices
\cite{1,2,3,4}, as well as because they are model objects for
various problems of micromagnetism \cite{5,6}.\,\,There are two
basic types of magnetic nanogranular (NG) composites: granules of a
ferromagnetic (FM) metal in a nonmagnetic insulating matrix or FM
granules in a nonmagnetic metal.\,\,In NG composites with a size of
FM granules less than the critical one for the single-domain state
\cite{7,8}, the remagnetization of a granule occurs by means of the
coherent rotation of the particle magnetization vector \cite{9}.
Below, we consider NG films of the type \textquotedblleft
single-domain FM metal granules in the insulating
matrix\textquotedblright\ without mentioning it every
\mbox{time.}\looseness=1

For external magnetic fields lower than the uniaxial anisotropy one,
the dependence of the magnetic energy of a granule on the
orientation of its magnetic moment looks like a double-well
potential \cite{10,11}.\,\,The magnetic moments of granules undergo
thermally induced transitions between those wells \cite{10,11}.

Describing the magnetic state of FM granules, an important concept
is the blocking temperature, $T_{\rm b}$~\cite{12}. Below $T_{\rm
b}$, the magnetic moment of the granule is \textquotedblleft
blocked\textquotedblright\ within a time interval of observation of
the system.\,\,In this case, the probabilities of its orientation
near each of two anisotropy energy minima do not correspond to the
thermodynamic equilibrium and depend on the prehistory.\,\,As a
result, the system demonstrates a hysteresis at the magnetization
reversal \cite{9,13}.\,\,Above $T_{\rm b}$, the thermodynamic
equilibrium for the populations of potential wells has enough time
to be established during the observation period, and the hysteresis
disappears \cite{12,14}.\,\,The ensemble of particles transits into
the superparamagnetic \mbox{state.}\looseness=1

However, as is shown in  works \cite{15,16,17}, the directions of
 particles' magnetic moments in each of two minima of the double-well
potential remain, to a great extent, localized close to these minima
up to temperatures (3$\div$4) $T_{\rm b}$, though the ratio of
``populations'' of those minima are practically equilibrium
\cite{10}.\,\,The corresponding mag\-ne\-tic-field and temperature
dependences of  the ensemble magnetization deviate from the Langevin
function \cite{15,16}. The latter corresponding to the equiprobable
orientation of the particles' magnetic moments into any direction in
the absence of an external field.\,\,The direction distribution of
magnetic moments in the film is not uniform in this case.

For a NG film with single-domain granules characterized by identical
magnitudes of anisotropy perpendicular to the film plane, the
emergence of a state with non-uniform distribution of magnetic
moment directions can also be connected with another factor.\,\,The
perpendicular anisotropy provides an energy minimum if the magnetic
moments of granules are oriented normally to the film plane, whereas
the anisotropy of the demagnetizing field, which is proportional to
the normal film magnetization, dictates to the film the easy-plane
anisotropy.\,\,In this case, the film magnetization in the field
directed normally to the film plane and less by magnitude than a
certain critical one, $H_{\mathrm{crit}}$, gives rise to the
emergence of an inhomogeneous state \cite{10}.\,\,In this state, the
magnetic moments of some FM granules are directed along the magnetic
field, and those of other granules in the opposite direction.\,\,The
demagnetization energy is lower, and the anisotropy energy has a
minimum.\,\,At the magnetization of a film with perpendicular
anisotropy of granules at an arbitrary angle $\theta _{\rm H}$, the
critical character of the transition between the homogeneous and
inhomogeneous states survives \cite{10}.\,\,The quantity
$H_{\mathrm{crit}}$ depends on $\theta _{\rm H}$ in this case and
it turn into zero at \mbox{$\theta _{\rm H}=90^{\circ}$.}

In the fields larger by magnitude than $H_{\mathrm{crit}}$, the
moments of all granules are oriented identically, and the whole
system is in the homogeneous state \cite{10}.\,\,However, at $\theta
_{\rm H}\neq0$, the moments are not oriented along the external
field: they are only tilted with respect to it until the
magnetization saturation is reached \cite{10}.

For the film as a whole, the field-induced transition between the
states with homogeneous and inhomogeneous magnetizations is a phase
transition of the order-disorder type \cite{10,18}.\,\,In the
equilibrium case and at the normal magnetization,
$H_{\mathrm{crit}}(\theta _{\rm H}=0)$ is equal to the largest
possible field of film demagnetization, $H_{d\mathrm{max}}=4\pi
M_{s}$.\,\,Here, $\theta _{\rm H}$ is the angle between the director
for the magnetic field direction and the film normal
($0\leqslant\theta _{\rm H}\leqslant\pi$), $M_{s}$ is the saturation
magnetization of the film as a whole ($M_{s}=f_{v}M_{s\_{\rm gr}}$),
$f_{v}$ is the volume fraction of the ferromagnetic material, of
which granules are made, and $M_{s\_{\rm gr}}$ is the saturation
magnetization of this material.\,\,The field $H_{d\mathrm{max}}$ is
achieved when the magnetization is saturated by a normally oriented
external field.

The state of a film with an inhomogeneous dis\-tri\-bu\-tion of
magnetic moment directions, which results from a competition between
the per\-pen\-di\-cu\-lar anisotropy and the demagnetization factor
ani\-so\-tro\-py, arises even at $T=0$, without thermally induced
transitions between the directions of magnetic moments in granules
that correspond to the minima of the doub\-le-well
potential.\,\,Just neg\-lec\-ting the thermally activated
transitions, the \textquotedblleft equi\-lib\-ri\-um
magnetization\textquotedblright\ of films with perpendicular
ani\-so\-tro\-py was considered in work \cite{10}.

An acceptable approximation for the consideration of the system in
this state is the model of two-level equilibrium system, which was
used, e.g., in the consideration of the equilibrium magnetization in
a nanogranular film with perpendicular anisotropy of particles
\cite{10}.\,\,In work \cite{10}, the behavior of the
\textquotedblleft equilibrium\textquotedblright\ (in the sense
indicated above) magnetization of the films with perpendicular
anisotropy of granules at various $\theta _{\rm H}$'s and the
dependences $H_{\mathrm{crit}}(\theta _{\rm H})$ for the isotropic
in-plane films with $f_{v}$ below the percolation threshold $f_{vp}$
were considered.\,\,The results of work \cite{10} were confirmed
experimentally \cite{19}.

In a number of cases, NG films with perpendicular anisotropy can
also have in-plane anisotropy.\,\,As far as we know, the regimes of
magnetic reversal for nanogranular films with perpendicular and
in-plane anisotropies have not been discussed in the literature.

As the volume fraction of granules in the film, $f_{v}$, grows,
there arise small clusters, and when $f_{v}$ reaches a value of
$f_{vp}$, there appears a large percolation cluster penetrating
throughout the film \cite{20}.\,\,However, even in this case, a
considerable fraction of FM granules remain isolated.\,\,At the
magnetic reversal, they have to demonstrate features typical of
films with perpendicular anisotropy.\,\,The researches of an NG film
with bi-phase magnetic content corresponding to the presence of
non-percolating granules and a percolation granular cluster in the
film were carried out in work \cite{21}. In this paper, the issues
concerning the resolution of contributions from those two groups of
granules to the magnetization reversal curves and the features in
the angular dependence of the coercive field in such film were
discussed.

In this work, we report the results of our additional researches of
a NG film Co/Al$_{2}$O$_{n}$ with 74.5~at.{\%} Co.\,\,This value
should correspond to $f_{v}\approx0.79$, which exceeds the
percolation threshold.\,\,We also considered the transition between
the homogeneous and inhomogeneous magnetization states of such
bi-phase film.\,\,Besides the perpendicular anisotropy, the granular
part of this film has the in-plane one as well \cite{21}.\,\,The
conditions for the critical transition in the granular part of the
film from the state with homogeneous orientation of magnetic moments
to the inhomogeneous one and their dependence on the direction of a
magnetizing field in the biaxially anisotropic film are
analyzed.

\section{Experimental Part}

\subsection{Specimens and measurement methods}

A batch of NG Co/Al$_{2}$O$_{n}$ film specimens about 5~$\mu\mathrm{m}$ in
thickness was fabricated by the ion-beam sputtering of composite targets located
at different ends of a long ($25\times1~\mathrm{cm}^{\mathrm{2}}$) sitall
substrate in the Ar atmosphere under a pressure of $3.2\times10^{-5}%
~\mathrm{Torr}$ \cite{18}.\,\,Such sputtering onto a long substrate
was used in order to provide the same conditions while fabricating a
set of films with various volume fractions of granules, $f_{v}$.
After the film sputtering, the substrate was cut across into
specimens: narrow strips with different Co contents.\,\,Therefore,
the specimens were fabricated under the conditions of the inclined
sputtering of components, especially the specimen with the minimum
and maximum Co contents.\,\,In this case, the preferential direction
in the film is the intersection of its plane with a plane formed by
atomic beams.\,\,This direction corresponds to the narrow side of
the specimen obtained after cutting the substrate.\,\,Let us
designate it as $Y$.\,\,The direction along the long side of a
specimen is designated as $X$, and the direction normal to the
specimen as $Z$.\,\,The dimensions of the examined sample of a film
amounted to 10 and 2~mm along the $X$ and $Y$ sides,
respectively.\,\,On the basis of the described procedure of film
fabrication (the geometry of sputtering and the relative arrangement
of the substrate and the targets), we may expect the manifestation
of anisotropic effects in the film \mbox{plane.}\looseness=1

The electric percolation threshold, which was determined for this
batch of films from the dependence of their electric resistance on
the Co content \cite{18}, was found to equal 61~at.\%~Co, which
should correspond to $f_{vp}=0.66$.\,\,Films from this batch with Co
contents from 45 to 75~at.\% were studied earlier in work \cite{22},
and the film-growth-induced perpendicular magnetic anisotropy was
found in them.\,\,It had the maximum magnitude near the percolation
threshold.\,\,In work \cite{23}, the  cross-section of a film from
this batch with a Co content of 56~at.\% was studied on a
transmission electron microscope.\,\,It was shown that the granules
with an average transverse size of about 4\textrm{~nm} are prolate
along the direction of their growth, i.e.\,\,along the normal of the
film. This fact explains the perpendicular anisotropy of the films,
as a form anisotropy.

As was mentioned above, we study an NG Co/Al$_{2}$O$_{n}$ film with
74.4~at.\%~Co ($f_{v}\approx0.79$), for which the percolation
threshold is considerably exceeded.\,\,In the previous research
\cite{21}, it was found that, in addition to the perpendicular
magnetic anisotropy, this film also has in-plane anisotropy with the
easy axis located in the film plane and oriented along the direction
$X$.\,\,The electron microscopy research of the transverse
cross-section of this film \cite{21} showed that the specimen
included both percolating clusters and non-percolating granules, and
the relative fraction of the latter dominated.\,\,The research of a
magnetic contrast on an atomic force microscope \cite{21} revealed
that, at room temperature, the specimen had a stripe domain
structure with the average domain width being about
1~$\mu\mathrm{m}$.\,\,The prevailing direction of the domain stripes
in the absence of an external field coincided with the axis $X$.
Hence, at room temperature, the examined film was in the
superferromagnetic state (a state with the ferromagnetic ordering of
superparamagnetic granules, which were separated from one another,
owing to the magnetic interaction between them).

The measurements of magnetization reversal curves were carried out
on a vibrating-sample magnetometer LDJ-9500 at room
temperature.\vspace*{-2mm}

\subsection{Results of measurements}

The families of magnetization reversal loops were registered at
various directions of the introduction of an external magnetic field
into the film, with $0\leqslant\theta _{\rm H}\leqslant$
$\leqslant90^{\circ}$ in the specimen planes $XZ$ ($\varphi_{\rm
H}=0$) and $YZ$ ($\varphi_{\rm H}=90^{\circ}$), which are shown
in Fig.~1.\,\,Hereafter, $\theta _{\rm H}$ ($0\leqslant\theta _{\rm H}%
\leqslant180^{\circ}$) is the angle between the direction of the
conditionally selected positive direction of the film normal and
external magnetic field director  (the slope angle of the axis,
along which the magnetic field $H$).\,\,Under such definition of the
field direction, the field itself can be considered as changing from
$+\infty$ to $-\infty$ at a constant $\theta _{\rm H}$ in the
spherical coordinate system used for calculations.\,\,The angle
$\theta _{\rm H}$ between the vector of the external magnetic field
strength $\mathbf{H}$ and the \textquotedblleft
positive\textquotedblright\ direction of the film normal depends in
this case on the field sign: $\theta _{\rm H}\ =\theta _{\rm
H}+[1-\mathrm{sign}(H)]\pi/2$, where $\mathrm{sign}(x)=1$, 0, and
$-1$ for $x>0$, $x=0$, and $x<0$, respectively.\,\,Accordingly, the
azimuthal angle $\varphi_{\rm H}$ for the direction of the magnetic
field vector should be defined as $\varphi_{\rm H}=\varphi_{\rm H}+[1-\mathrm{sign}%
(H)]\pi/2$.\,\,Here, $\varphi_{\rm H}$ is the azimuthal angle of the
director for the external magnetic field in the upper
hemisphere.\,\,The angle $\varphi_{\rm H}$ is reckoned from the
positive direction of the axis $X$.

The curves shown in Fig.~1 correspond to those expected for
ferromagnetic films with perpendicular and in-plane anisotropies and
a magnitude of $4\pi M_{s}$ for the maximum demagnetizing, which is
much larger than the fields of perpendicular ($H_{A\theta}$) and
in-plane ($H_{A\varphi}$) magnetic anisotropies of the film.

The hysteresis reveals itself unusually in the registered
curves.\,\,It is concentrated in a small part of magnetization
reversal curves, at low fields (see the insets in Fig.~1).\,\,In
work \cite{21}, it was shown that those hysteresis sections are the
manifestations of the contribution made by the percolating part of
the examined film to the magnetization reversal curve.\,\,The
percolating part feels the action of an internal field associated
with the dominating contribution of non-percolating granules to the
film magnetization.\,\,This internal field equals the vector sum of
the external and demagnetizing fields.\,\,The latter has only a
$Z$-component directed opposite to the $Z$-component of an external
field.\,\,The perpendicular anisotropy of FM granules results in
that the demagnetization almost completely compensates the
projection of an external field onto the axis $Z$ if the field is
lower than the critical one \cite{10}.\,\,At the same time, the
contribution of the percolating part of the film is responsible for
both the coercive field and its unusual angular dependence, which
was analyzed in work~\cite{21}.

\begin{figure}%
\vskip1mm
\includegraphics[width=7.8cm]{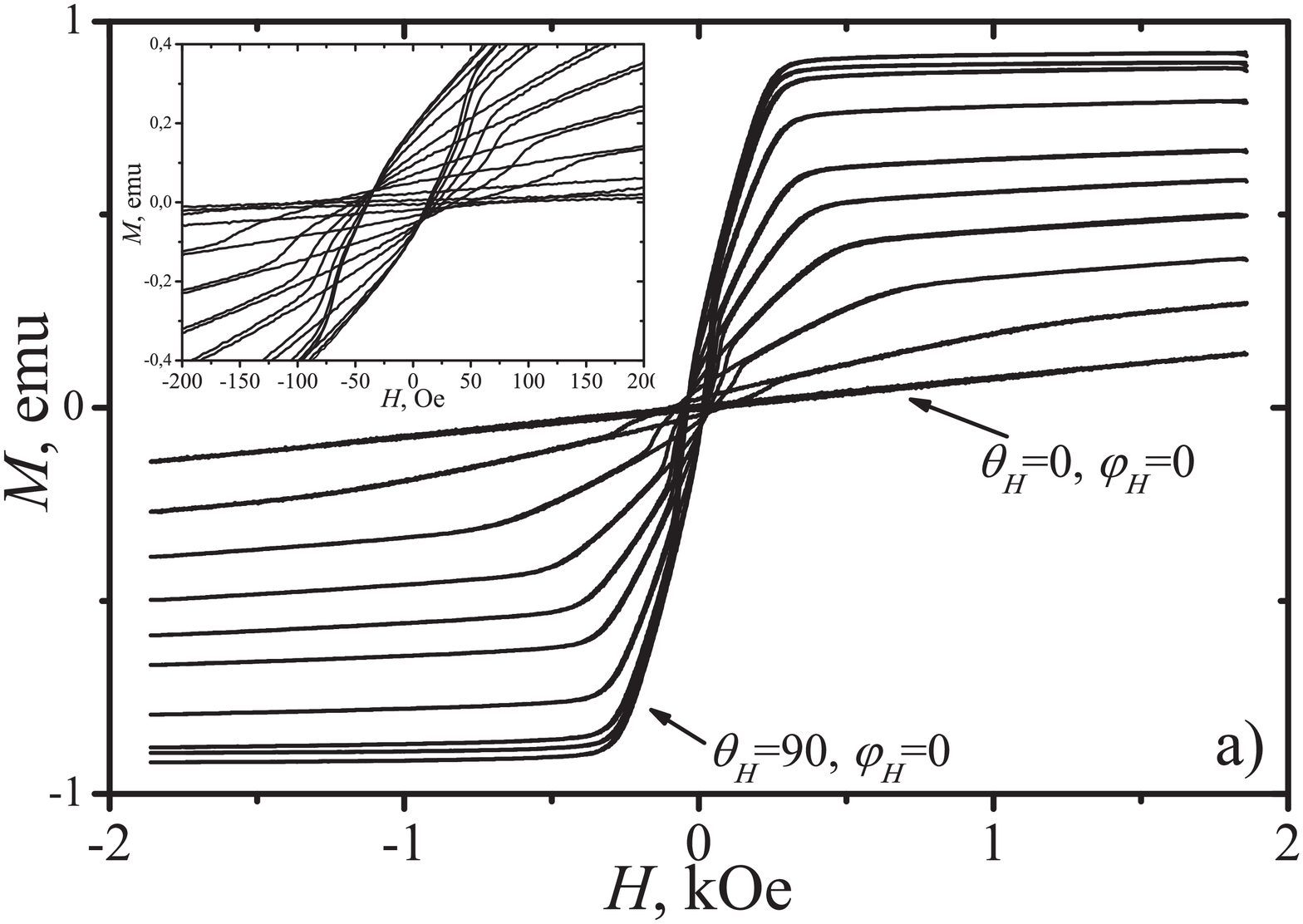}\\[2mm]
\includegraphics[width=7.8cm]{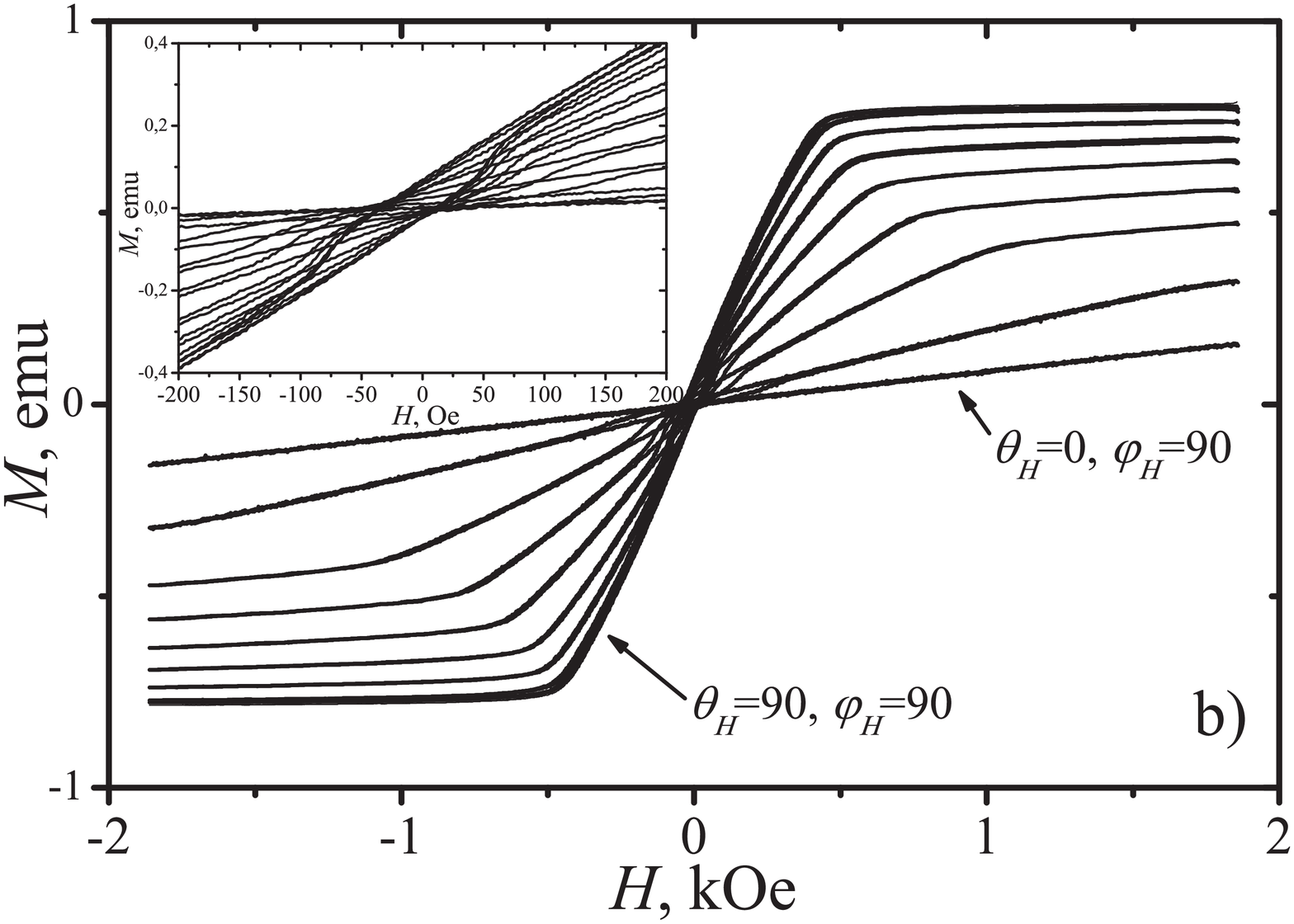}
\vskip-3mm\caption{Magnetization reversal loops for magnetic field
orientations in planes (\textit{a})$~XZ$, which corresponds to
$\varphi_{\rm H}=0$, and (\textit{b})~$YZ$, which corresponds to
$\varphi_{\rm H}=90^{\circ}$.\,\,Loops were measured with an angle
increment of $10^{\circ}$.\,\,The scaled-up fragments of central
loop sections are shown in the insets}
\end{figure}

\begin{figure}%
\vskip1mm
\includegraphics[width=7.6cm]{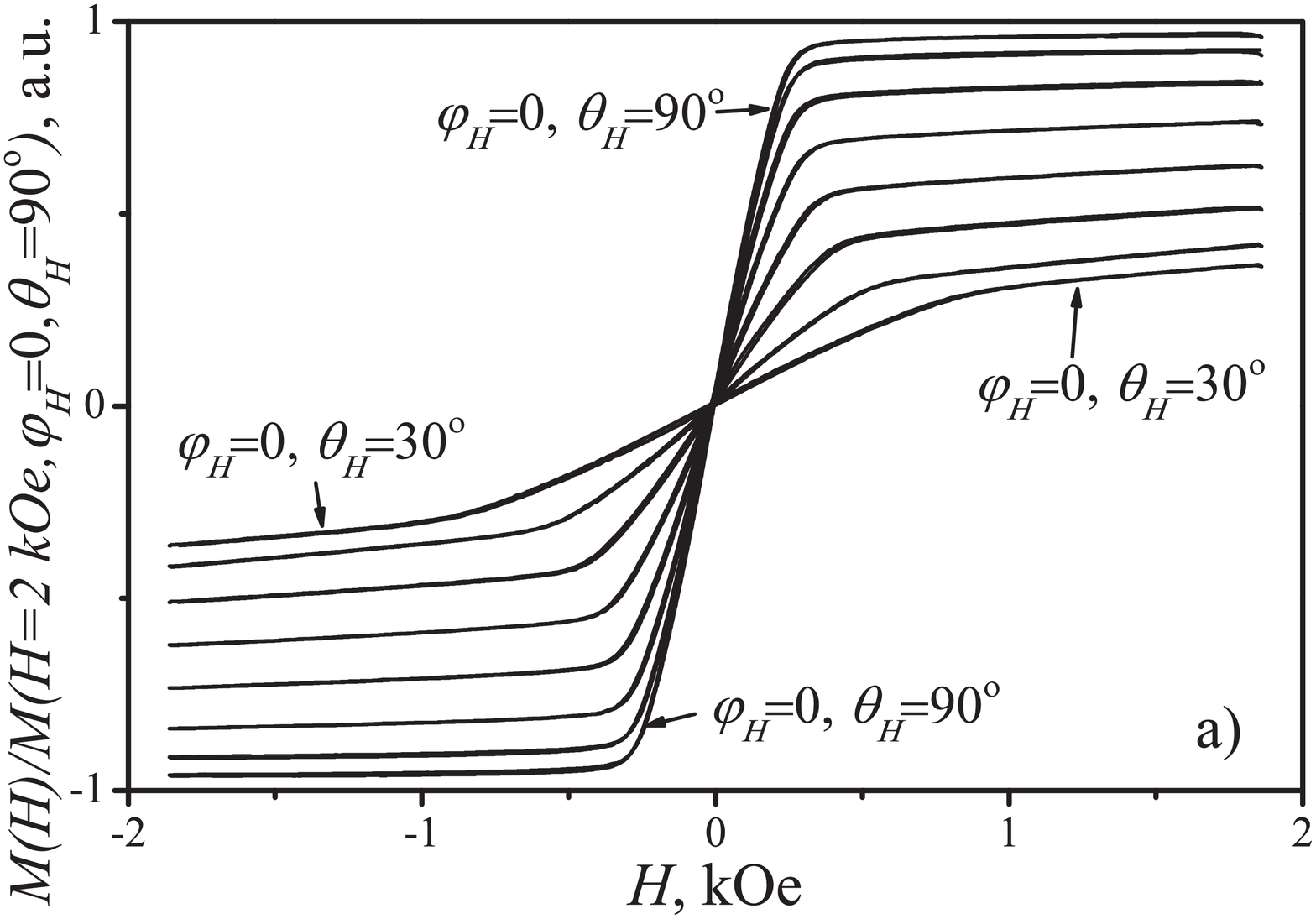}\\[2mm]
\includegraphics[width=7.6cm]{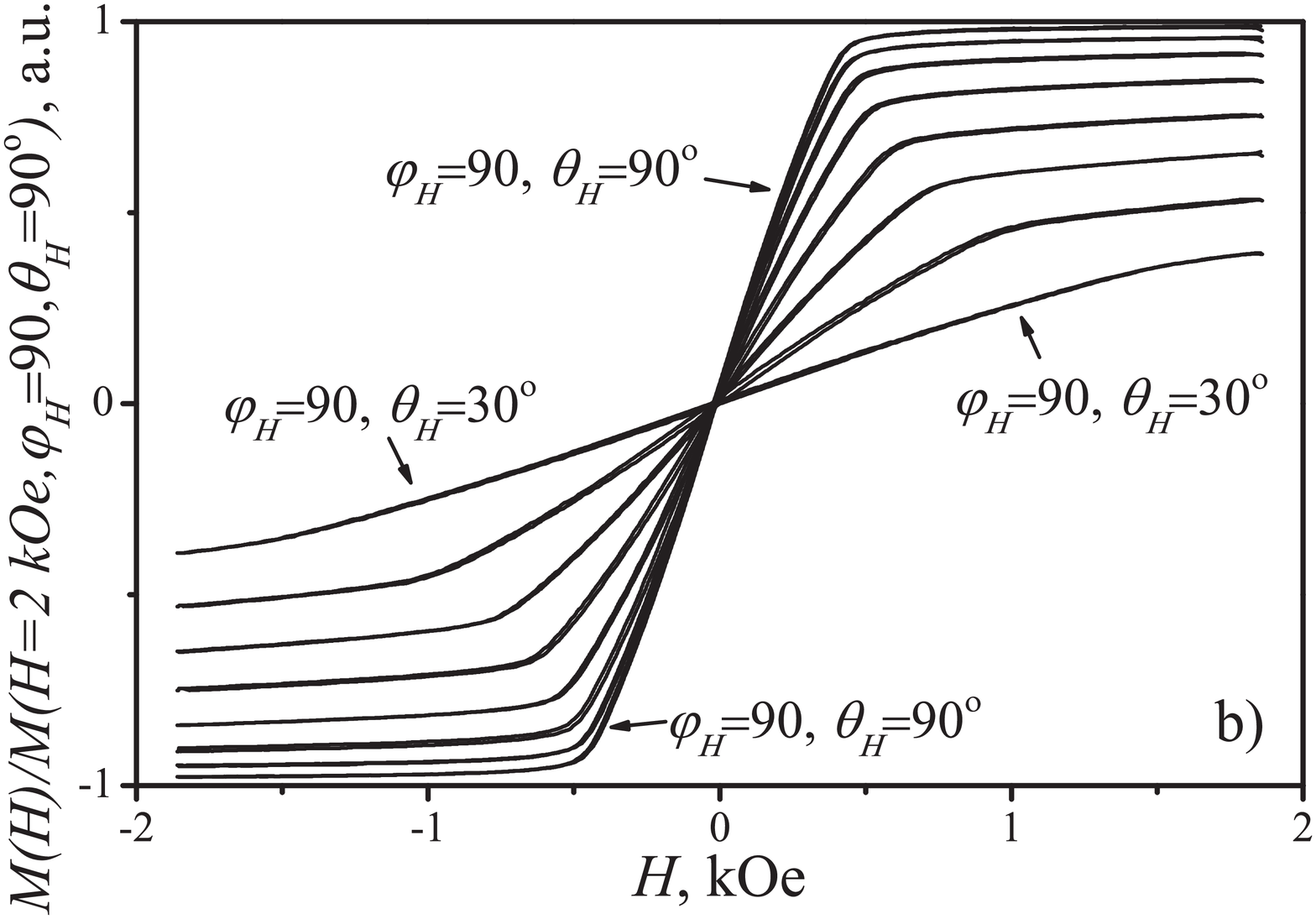}
\vskip-3mm\caption{Resolved granular contributions to the
magnetization reversal loop of NG Co/Al$_{2}$O$_{n}$ film for
various angles within the interval $30^{\circ }\leqslant\theta _{\rm
H}\leqslant90^{\circ}$ and $\mathbf{H}$ in the planes $XZ$
(\textit{a}) and $YZ$ (\textit{b})}
\end{figure}

\begin{figure}%
\vskip1mm
\includegraphics[width=\column]{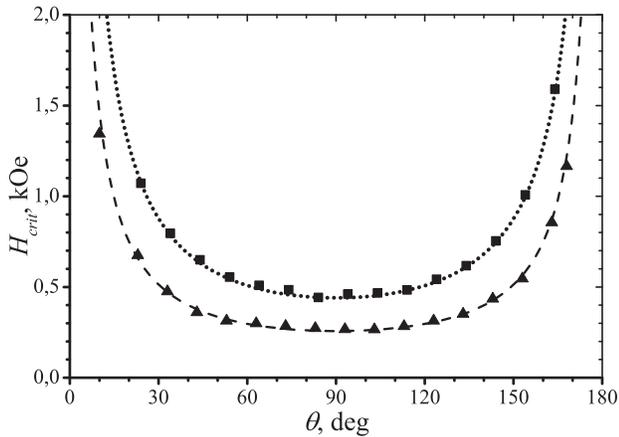}
\vskip-3mm\caption{Experimental dependences $H_{\rm crit}(\theta
_{\rm H})$ for the external field in the planes $YZ$ ($\varphi_{\rm
H}=90^{\circ}$, $\blacksquare$) and $XZ$ ($\varphi_{\rm
H}=0^{\circ}$, $\blacktriangle$).\,\,The curves correspond to the
approximation of experimental dependences at $\varphi_{\rm
H}=90^{\circ}$ and $0^{\circ}$ by expressions (17) and (18),
respectively }
\end{figure}

If we neglect the hysteresis in the narrow region of low fields,
then, in the field intervals between rather sharp transitions to the
saturation, the position of which depends on $\theta _{\rm H}$, and
$\varphi_{\rm H}$, the film magnetization is a linear function of
the field.\,\,Such a behavior almost corresponds to the model of
equilibrium magnetization reversal in a tilted magnetic field for
the film with perpendicular anisotropy, which was considered in
works \cite{10,19}.\,\,The only difference consists in that the
 fields of the sharp inflection to saturation in the dependences $M_{\rm H}(\theta _{\rm H})$
are different for the field direction planes $XZ$ and $YZ$.\,\,In
nanogranular films with perpendicular anisotropy, those fields of
the sharp inflection to saturation are critical, i.e.\,\,the
equilibrium inhomogeneous ensemble of granules transforms in them
into the equilibrium homogeneous one \cite{10}.

In work \cite{21}, we proposed a method to resolve the obtained
magnetization reversal curves into components given by the
\textquotedblleft granular\textquotedblright\ and \textquotedblleft
percolating\textquotedblright\ parts of the film with the use of the
derivatives of magnetization reversal curves.\,\,Since the aim of
the present work is to analyze the dependences of critical fields,
at which the granular part of the film transits from the equilibrium
inhomogeneous state of the ensemble of granules into the equilibrium
homogeneous one, only the granular contribution to the magnetization
reversal curves resolved using the method proposed in work \cite{21}
will be considered in what follows.

The normalized magnetization reversal loops for this contribution
are shown in Fig.~2,~\textit{a,~b} for the field direction planes XZ
and YZ, correspondingly.\,\,It should be noted that, at $\theta
_{\rm H}<30^{\circ}$, the sharp inflections in the curves shift
toward larger fields, and the approach to the saturation level after
it becomes very slow, which is associated with a large value of
$4\pi M_{s}$ for the film concerned.\,\,Therefore, the curves for
those angles were not plotted in Fig.~2.\,\,In both \textit{a} and
\textit{b} cases, all curves demonstrate a behavior typical of NG
films with perpendicular anisotropy at \textquotedblleft
equilibrium\textquotedblright\ (in the sense discussed in
Introduction): an almost linear $M(H)$-dependence, free of
hysteresis, in low fields; then a sharp inflection and a smooth
asymptotic approach to the saturation level at high fields.\,\,Both
the slope angle of the linear section and the position of the
critical transition point (the inflection point) have their own
angular dependences.\,\,It should also be noted that the slope angle
of the linear section at low fields in the dependence $M(H)$ in
Fig.~2,~\textit{a} differs insignificantly from that in
Fig.~2,~\textit{b}.\,\,Additionally, the dashed curves in Fig.~3
approximate the dependences $H_{\mathrm{crit}}(\theta _{\rm H})$ by
expressions, which are derived in the next section.

\section{Model}

As was mentioned above, in work \cite{10}, the magnetization
reversal in an arbitrarily directed external field and the critical
transition from the state with homogeneous magnetization to the
state with different orientations of magnetic moments in two
subensembles of FM granules were considered for a NG film, which is
isotropic in its plane, but with perpendicular anisotropy, and with
the volume fraction occupied by granules, $f_{\mathrm{gr}}$, below
the percolation threshold.\,\,That is, the volume fraction of a FM
material in the film, $f_{v}$, was considered to equal $f_{\rm gr}$.
From the discussion of experimental data, it follows that two
factors in the case considered in this work are different.\,\,First,
the examined film is above the percolation threshold, i.e.
$f_{v}>f_{vp}$ for it, so that it consists of \textquotedblleft
granular\textquotedblright\ $f_{\mathrm{gr}}<f_{v}$ and
\textquotedblleft percolating\textquotedblright\ parts.\,\,Second,
besides the perpendicular anisotropy, this film is also
characterized by the in-plane one.\,\,Let us consider how those
differences can be taken into account on the whole in the framework
of the approach used in works \cite{10,21}.

\subsection{Role of the overpercolating\\ state of a film and its account}

\label{sec3.1}

In order to consider the overpercolating state of a film, let us
introduce the relative volume of a FM metal in the composite, which
consists of the separate granules, $f_{\rm gr}$, and the relative
volume of this material in the percolation cluster, $f_{\rm
per}$.\,\,Then $f_{v}=f_{\rm gr}+f_{\rm per}$.

In this subsection, for simplification, we neglect the in-plane
anisotropy.\,\,Since we consider the equilibrium (in the sense
described in Introduction) state of the film, we may confine the
consideration to the interval of magnetic fields $H>0$ and select
the reference point for the angle $\varphi$ so that $\varphi_{\rm
H}=0$. Then, at a field larger than the critical one, when the
magnetic moments of all granules are oriented identically, the
magnetic energy density in the two-component system consisting of
granules with
perpendicular anisotropy and the percolating part looks like%
\[ U=-f_{\rm gr} Kj^2\cos ^2\theta _{\rm gr} \,+\]\vspace*{-7mm}
\[+\,\frac{1}{2}N_{zz} ( {f_{\rm gr} j\cos
\theta _{\rm gr} +f_{\rm per} j\cos \theta _{\rm per} } )^2\,-
\]\vspace*{-7mm}
\[
 -\,H( \cos \theta _{\rm H} ( {f_{\rm gr} j\cos \theta _{\rm gr} +f_{\rm per} j\cos
\theta _{\rm per} } )\,+\]\vspace*{-7mm}
\begin{equation}
\label{eq1} +\,\sin \theta _{\rm H} ( {f_{\rm gr} j\sin \theta _{\rm
gr} +f_{\rm per} j\sin \theta _{\rm per} } ) ).
\end{equation}
Here, the first term describes the contribution to the energy from
the uniaxial anisotropy of the non-percolating part of granules
($K>0$ is the anisotropy constant for the non-percolating part of
granules), the second term is the demagnetization energy ($N_{zz}$
is the demagnetizing factor), and the third term is the Zeeman
contribution.\,\,The angles of the magnetic moment directions of
granules, $\theta _{\rm gr}$, and the percolating part, $\theta_{\rm
per}$, are not identical.\,\,The filling factors of the film with
granules, $f_{\rm gr}$, and with the percolating part, $f_{\rm
per}$, were also taken into account.\,\,The absolute values of their
own magnetizations, $j$, were adopted to be equal.

The equation for the extremes of the energy density for a two-component
system in the state where the magnetic moments of granules are directed
identically can be obtained by differentiating Eq.~(\ref{eq1}) with respect to
the angles and equating the derivatives to zero,
\[
 \frac{\partial U}{\partial \theta _{\rm gr} }=2f_{\rm gr} Kj^2\cos \theta _{\rm gr}
\sin \theta _{\rm gr} \,-\]\vspace*{-7mm}
\[-\,N_{zz} f_{\rm gr} j^2( {f_{\rm gr}
\cos \theta _{\rm gr} +f_{\rm per} \cos \theta _{\rm per} } )\sin
\theta _{\rm gr} \,-
\]\vspace*{-7mm}
\begin{equation}
\label{eq2}
 -\,Hf_{\rm gr} j( {-\cos \theta _{\rm H} \sin \theta _{\rm gr} +\sin \theta _{\rm H} \cos
\theta _{\rm gr} } )=0 ,
\end{equation}\vspace*{-7mm}
\[
 \frac{\partial U}{\partial \theta _{\rm per} }\!=\!-\!N_{zz} f_{\rm per} j^2(
f_{\rm gr} \cos \theta _{\rm gr} \!+\!f_{\rm per} \cos \theta _{\rm
per} )\sin \theta _{\rm per} \,- \]\vspace*{-7mm}
\begin{equation}
\label{eq3}
 -\,Hf_{\rm per} j\left( {-\cos \theta _{\rm H} \sin \theta _{\rm per} +\sin \theta _{\rm H} \cos
\theta _{\rm per} } \right)=0.
\end{equation}
Let us consider a state of the system, in which the demagnetizing
field is compensated by the $Z$-projection of the external magnetic
field.\,\,This is the lowest value of external field, at which the
magnetic moments of all granules still remain oriented identically.
Let us define it as \textquotedblleft critical\textquotedblright:
\begin{equation}
\label{eq4} N_{zz} j(f_{\rm gr} \cos\theta _{\rm gr} +f_{\rm per}
\cos\theta _{\rm per} )-H_{\rm crit} \cos\theta _{\rm H} =0.
\end{equation}
Substituting Eq.~(\ref{eq4}) into Eqs.~(\ref{eq2}) and (\ref{eq3})
allows us to find the external field for this critical point and the
corresponding magnetization orientations $\theta_{\rm gr}^{\rm
(cr)}$ and $\theta_{\rm per}^{\rm (cr)}$.\,\,Note that earlier, in
work \cite{10}, the critical field for the film with perpendicular
anisotropy of granules with $f_{v}<f_{vp}$ was defined as the
limiting case for the state, in which the granules form two
subensembles.

From Eqs.~(\ref{eq2}), (\ref{eq3}), and (\ref{eq4}), we obtain that the
following conditions are satisfied at the critical point:%
\begin{equation}
\begin{array}{l}
N_{zz} = f_{\rm gr} j \cos \theta_{\rm gr}^{\rm (cr)} -H_{\rm cr}
\cos \theta_{\rm H} =0, \\[2mm]
2 f_{\rm gr}Kj\sin \theta_{\rm gr}^{\rm (cr)} -H_{\rm cr} \sin
\theta_{\rm H} =0,\quad \cos \theta_{\rm per}^{\rm (cr)}=0.
\end{array}\!\!\!\!\!\!\!\label{5}
\end{equation}
While comparing Eqs.~(\ref{5}) with the corresponding equations in
work \cite{10}, one can see that the former give the same values for
the critical field in the case $f_{v}<f_{vp}$; now, the equations
include the relative volume of non-percolating granules, $f_{\rm
gr},$ rather than the whole relative volume of a ferromagnetic
substance in the NG film.\,\,In the non-percolating film, which was
considered in work \cite{10}, those two quantities were identical.

The solution of Eqs.~(\ref{eq2}) and (\ref{eq3}) makes it possible
to find the required dependences for the projection of the magnetic
moment of a percolating film onto the external field if the latter
exceeds the critical value ($\left\vert \mathbf{H}\right\vert
>H_{\rm crit}$).\,\,Below the critical point ($\left\vert
\mathbf{H}\right\vert <H_{\rm crit}$), the magnetic moments of
 granules in the film are not oriented identically, which
diminishes the positive demagnetization energy, but keeps the gain
provided by the anisotropy energy, as was described in work~\cite{10}.

\subsection{Biaxial anisotropy}

The description of the critical transition induced by a magnetic
field in the NG film with perpendicular anisotropy of granules
between the state with uniformly oriented magnetic moments of all
granules and the state with their non-uniform orientation, which was
used in work \cite{10}, was based on the model that, in the fields
with $\left\vert \mathbf{H}\right\vert <H_{\rm crit}$, there exist
two subensembles of granules with different average orientations of
magnetic moments in the film.\,\,As a result, the positive energy of
demagnetization becomes lower, but the gain in the negative energy
of perpendicular anisotropy is preserved.\,\,The relative numbers of
granules related to either of the subensembles are denoted as
$p_{1}$ and $p_{2}$ ($p_{1}+p_{2}=1$).\,\,The states with
homogeneous magnetization at $\left\vert \mathbf{H}\right\vert
>H_{\rm crit}$ correspond to the combination $p_{1}=1$ and $p_{2}=0$
(at $H>H_{\rm crit}$) or the combination $p_{1}=0$ and $p_{2}=1$ (at
$-H<-H_{\rm crit}$).\looseness=1

Let us use a similar approach in the case where the granular part of
a film has both perpendicular and in-plane anisotropies.\,\,On the
basis of the consideration in Section~\ref{sec3.1}, we analyze only
the contribution of the granular part, taking into account that the
relative volume
occupied by non-percolating granules in the film equals $f_{\rm gr}=f_{v}-f_{\rm per}%
$.\,\,We also confine the consideration to the case $H>0$.\,\,Taking
the indicated factors into account, the equation for the total
magnetic energy in the
granular part reads%
\[ U_{\rm tot} =f_{\rm gr} \{-K_\theta (p_1\cos^{2}\theta _1 +p_2
\cos^{2}\theta _2 )\,+\]\vspace*{-7mm}
\[+\,K_\varphi (p_1 {\cos}^{2}\varphi _{1} {\sin}^{2}\theta _1 +p_2 {\cos}^{2}\varphi _{2} \sin^{2}\theta
_2)\,+\]\vspace*{-7mm}
\[ +\,\frac{{1}}{{2}}N_{zz} f_{\rm gr} j^{2}(p_1
{ \cos}\theta _1 +p_2 { \cos}\theta _2)^{2} \,-\]\vspace*{-7mm}
\[-\,Hj[(p_1
{\cos}\theta _1 +p_2 {\cos}\theta _2 ) \cos\theta _{\rm H}\,+
\]\vspace*{-7mm}
\[
 +\,(p_1 \cos(\varphi _1 -\varphi _{\rm H} )\sin \theta _1 \,+\]\vspace*{-7mm}
\begin{equation}
\label{eq5}
 +\,p_2\cos(\varphi _2 -\varphi _{\rm H} ) \sin\theta_2 ) \sin\theta_{\rm H} ]\}.
\end{equation}
Here, $U_{\rm tot}$ is the total energy density in the granular part, $j$ the magnetic moment of a granule, $K_{\theta}$
the perpendicular
anisotropy constant, $K_{\varphi}$ the in-plane anisotropy constant, $\theta_{1}%
$, $\theta_{2}$, $\varphi_{1}$, and $\varphi_{2}$ are the angles of
magnetic moment directions of granules in the subensembles measured
in spherical coordinates, and $\theta _{\rm H}$ and $\varphi_{\rm
H}$ are the director tilt angles for the external magnetic field
$H$.\,\,For simplification, let us introduce the quantity $\Delta
p=p_{1}-p_{2}$.\,\,Then, $p_{1}=\frac{1+\Delta p}{2}$ and
$p_{2}=\frac{1-\Delta p}{2}$.\,\,The extremes of $U_{\rm tot}$ with
respect to the parameters $\theta_{1}$, $\theta_{2}$, $\varphi_{1}$,
$\varphi_{2},$ and $\Delta p$ can be found by equating the
corresponding derivatives to zero:
\[
 \frac{\partial U_{\rm tot} }{\partial \theta _{1} }=f_{\rm gr} \{(
1 +\Delta {p} )K_\theta  \cos\theta _{1}  \sin\theta _{1}\,
+\]\vspace*{-7mm}
\[+\,(1+\Delta p )K_\varphi \cos\varphi _{1}  \cos^{2}\varphi _{1}
\sin\theta _{1}\,- \]\vspace*{-7mm}
\[-\,\frac{1}{4}f_{\rm gr} j^{2}Nzz(( 1+\Delta p) \cos\theta _{1} \,+\]\vspace*{-7mm}
\[+\,( 1-\Delta p ) \cos\theta _{2})( 1+\Delta p) \sin\theta _{1}\,-\]\vspace*{-7mm}
\[
 -\,\frac{{1}}{{4}}H j( 1+\Delta p
)( \cos\theta _{1}  \cos( \theta _{1} -\theta _{\rm H}  )\sin\theta
_{\rm H}\, -\]\vspace*{-7mm}
\begin{equation}
\label{eq6} -\, \cos\theta _{\rm H}  \sin \theta _{1} )\}=0 ,
\end{equation}\vspace*{-7mm}
\[
 \frac{\partial U_{\rm tot} }{\partial \theta _2 }=f_{\rm gr} \{(
1-\Delta p)K_\theta \cos\theta _{2}  \sin\theta _{2}
\,+\]\vspace*{-7mm}
\[+\,(1-\Delta p )K_\varphi j \cos\theta _{2}\cos^{2}\varphi _{2}
\sin\theta _{2}\,-\]\vspace*{-7mm}
\[ -\,\frac{1}{4}( 1-\Delta p)f_{\rm gr} j^{2}N_{zz} ( ( 1+\Delta p )\cos \theta _{1}\, +\]\vspace*{-7mm}
\[+\,( 1-\Delta p )\cos\theta _{2} )\sin\theta _{2} \,+\]\vspace*{-7mm}
\begin{equation}
\label{eq7}
 +\frac{{1}}{{2}}(1 -\Delta p)Hj\cos\theta _{\rm H} \sin\theta _{2}
 \}={0},
\end{equation}\vspace*{-5mm}
\[ \frac{\partial U_{\rm tot} }{\partial \varphi _1 }=f_{\rm gr} \{-(
{{1}+\Delta p} )K_\varphi j\cos\varphi _{1} \sin^{2}\theta _{1}
\sin\varphi _{1} \,+\]\vspace*{-5mm}
\begin{equation}
\label{eq8} +\,\frac{1}{2}( 1+\Delta p )H j \sin\theta _{1}
\sin\theta _{\rm H} \sin( \varphi _{1} -\varphi _{\rm H})\}={0},
\end{equation}
\[ \frac{\partial U_{\rm tot} }{\partial \varphi _2 }=f_{\rm gr} \{-(1-\Delta p)K_\varphi \cos\varphi _{2} \sin^{2}\theta
_{2} \sin\varphi _{2}\, +\]\vspace*{-6mm}
\begin{equation}
\label{eq9} +\,\frac{{1}}{{2}} (1-\Delta p) H j\sin^{2}\theta _{\rm
H} \sin( {\varphi _2 -\varphi _{\rm H} } )\}={0},
\end{equation}\vspace*{-6mm}
\[
 \frac{\partial U_{\rm tot} }{\partial \Delta p}=f_{\rm gr}\Bigl \{\!-\frac{{1}}{{2}}K_\theta ( \cos^{2}\theta _{1} -\cos^{2}\theta _{2}
)\,+\]\vspace*{-7mm}
\[+\,\frac{1}{2}K_\varphi (
\cos^{2}\varphi _{1} \sin^{2}\theta _{1} -\cos^{2}\varphi _{2}
\sin^{2}\theta _{2} )\,+ \]\vspace*{-7mm}
\[ +\,\frac{1}{4}f_{\rm gr} j^{2}N_{zz} ( \cos\theta _{1} -\cos\theta _{2}
)( ( 1+\Delta p)\,\times \]\vspace*{-7mm}
\[\times\,\cos\theta _{1} +( 1-\Delta
p)\cos\theta _{2})\,-\]\vspace*{-7mm}
\[-\,\frac{{1}}{{2}}H j( (
\cos\theta _{1} -\cos\theta _{2})\cos\theta _{\rm H}+\sin\theta
_{\rm H} ( \cos( \varphi _{1}-\varphi _{\rm H}
)\,\times\]\vspace*{-7mm}
\begin{equation}
\label{eq10}  \times\, \sin\theta _{1} -\cos( \varphi _{2} -\varphi
_{\rm H} )\sin\theta _{\rm H} ) )\!\Bigr\}={0}.
\end{equation}
From the system of equations (\ref{eq6}), (\ref{eq7}), and (\ref{eq10}), it
follows that, in the equilibrium state with $\left\vert \Delta p\right\vert
\neq1$,%
\begin{equation}
\theta _1 +\theta _2 =180^{\circ }.\label{12}
\end{equation}

The solution of the system of equations (\ref{eq6})--(\ref{eq10})
for an arbitrary angle $\varphi_{\rm H}$ is rather complicated even
in the numerical form.\,\,Therefore, let us consider only two cases:
$\varphi_{\rm H}=0$ and 90$^{\circ}$.\,\,From Eqs.~(\ref{eq8}) and
(\ref{eq9}), one can see that, in those cases, the condition
$\varphi_{1}=\varphi_{2}=\varphi_{\rm H}$ is satisfied.\,\,For
convenience, let us introduce the anisotropy fields and the maximum
possible demagnetizing field created by the granular part of the
film ($\Delta
p=0$ or 1 if $\theta _{\rm H}=0$), by using the substitutions $H_{a\theta}%
=\frac{2K_{\theta}}{j}$, $H_{a\varphi}=\frac{2K_{\varphi}}{j}$, and
$H_{d}^{\max}=j\,f_{\rm gr}N_{zz}$.

In the fields below the critical one, the state of equilibrium
inhomogeneous magnetization is realized.\,\,With regard for
expression (\ref{12}), let us solve the system of equations
(\ref{eq2}) and (\ref{eq5}) in the cases where $\varphi_{\rm H}=0$
and 90$^{\circ}$.\,\,For the system in the equilibrium inhomogeneous
state, the parameters $\Delta p$ and $\theta_{1}$ look as follows:

if $\varphi_{\rm H}=0$,\vspace*{-2mm}
\begin{equation}
\label{eq11} \theta _1 (\varphi _{\rm H} =0)={\rm arcsin}\left(\!
{\frac{H\sin\theta _{\rm H} }{H_{a\theta } -H_{a\varphi } }}\!
\right)\!,
\end{equation}\vspace*{-5mm}
\begin{equation}
\label{eq12} \Delta p(\varphi _{\rm H} =0)=\frac{H{ \cos}\theta
_{\rm H} }{H_d^{\max} \sqrt {{1-}\frac{H^{2}\sin^{2}\theta _{\rm H}
}{(H_{a\theta } -H_{a\varphi } )^{2}}} },
\end{equation}

if $\varphi_{\rm H}=90^{\circ}$,\vspace*{-2mm}
\begin{equation}
\label{eq13} \theta _1 (\varphi _{\rm H} =90^{\circ })={\rm
arcsin}\left(\! {\frac{H \sin\theta _{\rm H} }{H_{a\theta } }}\!
\right)\!
\end{equation}\vspace*{-7mm}
\begin{equation}
\label{eq14} \Delta p(\varphi _{\rm H} =90^{\circ })=\frac{H{
\cos}\theta _{\rm H} }{H_d^{\max} \sqrt
{{1-}\frac{H^{2}{\sin}^{2}\theta _{\rm H} }{H_{a\theta } ^{2}}} }.
\end{equation}

\noindent At the critical transition point, $\left\vert \Delta
p\right\vert $ becomes equal to 1, and two subensembles merge into
one.\,\,From expressions (\ref{eq12}) and (\ref{eq14}), we find the
critical field magnitude:

if $\varphi_{\rm H}=0$,%
\[ H_{\rm crit} (\varphi _{\rm H} =0)=\]\vspace*{-7mm}
\begin{equation}
\label{eq15} =\frac{ (H_{a\theta } -H_{a\varphi } ) H_d^{\max}
}{\sqrt { (H_{a\theta } -H_{a\varphi } )^{2} \cos^{2}\theta _{\rm H}
+(H_d^{\max })^2 {\sin}^{2}\theta _{\rm H} } },
\end{equation}

if $\varphi_{\rm H}=90^{\circ}$,%
\begin{equation}
\label{eq16} H_{\rm crit} (\varphi _{\rm H} =90^{\circ
})=\frac{H_{a\theta } H_d^{\max } }{\sqrt {H_{a\theta }^2 {
\cos}^{2}\theta _{\rm H} +(H_d^{\max })^2 { \sin}^{2}\theta _{\rm H}
} }.
\end{equation}

\noindent In the case where $H\geqslant H_{\rm crit}$, the system of
superparamagnetic granules transits into the homogeneously
magnetized state, which is characterized by the parameters
$\left\vert \Delta p\right\vert =1$ and
$\theta_{1}=\theta_{2}=\theta m$.\,\,However, bearing in mind that
the system also contains the percolating part, the contribution of
the latter to the demagnetizing field has to be taken into
account.\,\,Neglecting this contribution in the case of a
non-percolating film with perpendicular and in-plane anisotropies,
the derivative of $U_{\rm tot}$ with respect to $\theta_{\rm crit}$
for the homogeneously magnetized system looks as follows:

if $\varphi_{\rm H}=0$,%
\[ \frac{\partial U_{\rm tot} }{\partial \theta _m }=Hj\cos\theta _{\rm H}
\sin\theta _m -j\cos\theta _m  (H\sin\theta _{\rm H}
\,+\]\vspace*{-7mm}
\begin{equation}
\label{eq17} +\,(-H_{a\theta } +H_{a\varphi } +H_d^{\max} )
\sin\theta _m ),
\end{equation}

if $\varphi_{\rm H}=90^{\circ}$,%
\[
\frac{\partial U_{\rm tot} }{\partial \theta _m }=-j(H \sin(\theta
_{\rm H} -\theta _m )\,+\]\vspace*{-7mm}
\begin{equation}
+\,H_d^{\max } \cos\theta _m \sin\theta _m )+\frac{j}{2}H_{a\theta }
\sin(2\theta _m ).\label{20}
\end{equation}

One can obtain the equations for the equilibrium values of
$\theta_m$ by equating  (19) and (20) to zero.

\section{Comparison\\ of the Model with the Experiment}

\begin{figure}%
\vskip1mm
\includegraphics[width=\column]{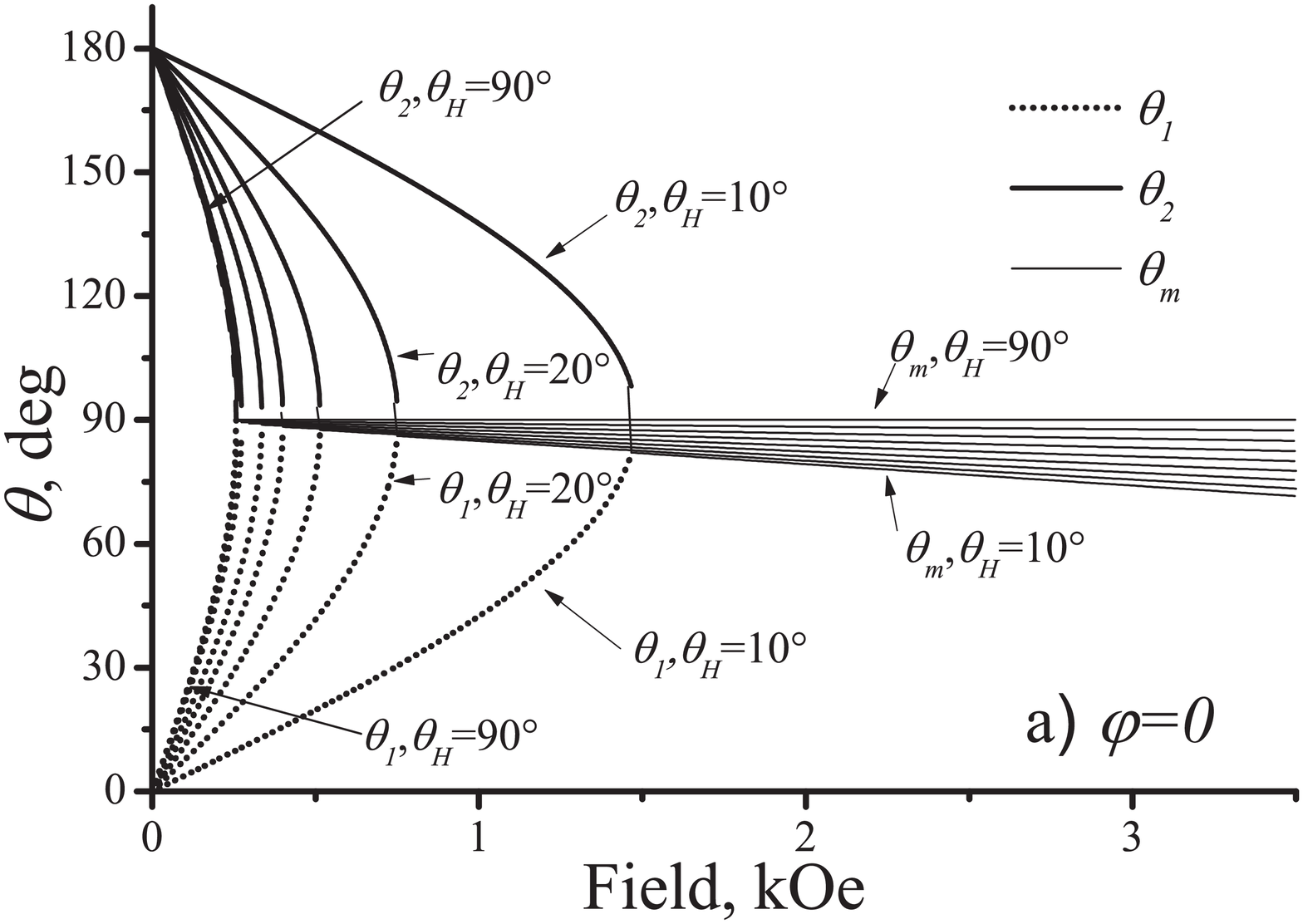}\\[2mm]
\includegraphics[width=\column]{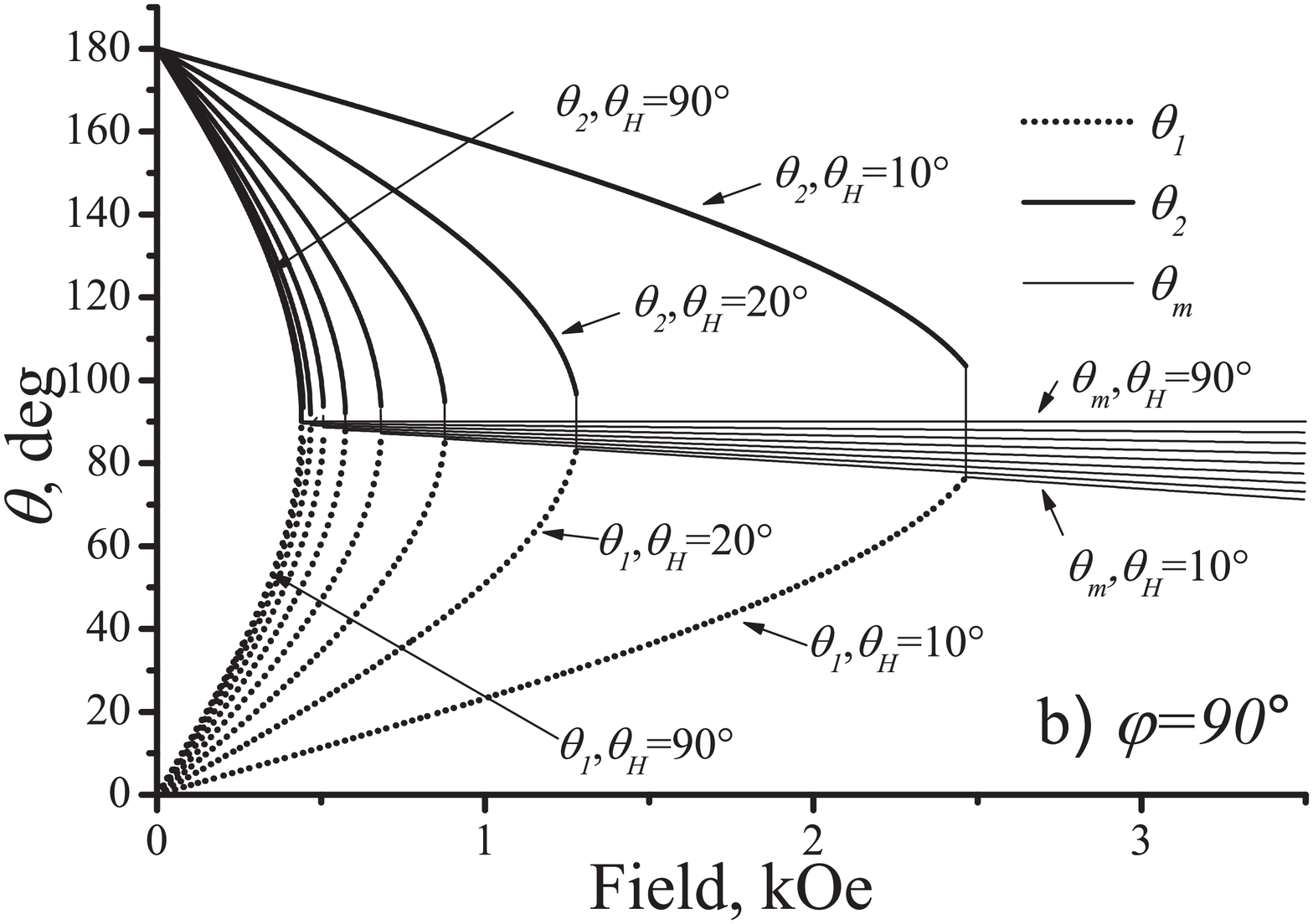}
\vskip-3mm\caption{Dependences of magnetization orientations for the
subensembles of NG particles on the external magnetic field for
various $\theta _{\rm H}$: (\textit{a})~$\varphi_{\rm H}=0$ and
(\textit{b})~$\varphi_{\rm H}=90^{\circ}$}
\end{figure}

\begin{figure}%
\vskip1mm
\includegraphics[width=\column]{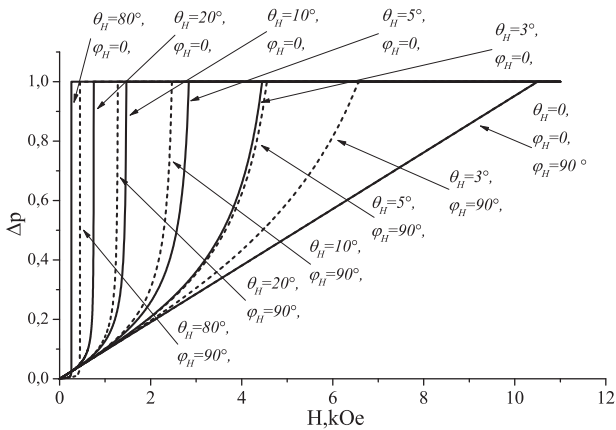}
\vskip-3mm\caption{Population difference $\Delta p$ at the
magnetization in various directions within the interval
$0^{\circ}\leqslant\theta _{\rm H}\leqslant 90^{\circ}$ (an
increment of 10$^{\circ}$) for two planes of the magnetic field
orientations: with $\varphi _{\rm H}=0$ (solid curves) and with
$\varphi_{\rm H}=90^{\circ}$ (dashed curves) }
\end{figure}

By approximating the experimental data obtained for the dependence $H_{\rm crit}%
(\theta _{\rm H})$ in the cases where the external field is located
in the planes $YZ$ ($\varphi_{\rm H}=90^{\circ}$) and $XZ$
($\varphi_{\rm H}=0$) (see Fig.~3), we determined the values
$H_{d}^{\max}=10.5\,\pm\,0.5$~kOe, $H_{a\theta }=0.44\,\pm$
$\pm\,0.02$~kOe, and $H_{a\varphi}=0.183\,\pm\,0.02$~kOe.\,\,The
equilibrium dependences for $\theta_{1}$, $\theta_{2}$ and
$\theta_{m}$, which are described by Eqs.~(\ref{eq11}) and
(\ref{eq13}), were calculated for the indicated values; they are
depicted in Fig.~4 for the external magnetic field orientations in
the planes that contain the normal to the film and the easy
(Fig.~4,~\textit{a}) or hard (Fig.~4,~\textit{b}) axis of the
in-plane anisotropy.\,\,Since, according to the estimations carried
out in work \cite{21}, the contribution made by the percolating part
of the film to its magnetization in the saturated state does not
exceed 20\%, the quantity $\theta_{m}$ was calculated by equating
derivatives (\ref{eq17}) and (\ref{20}) to zero and neglecting this
contribution.\,\,In the inhomogeneous state, the moments of both
subensembles deflect toward the equilibrium direction almost
linearly, with a drastic ultimate transition into the homogeneous
state; afterward, the direction of the magnetic moments in the
single, united ensemble with $\Delta p=1$ asymptotically approaches
the direction of $\mathbf{H}$.\,\,The difference from the
corresponding solutions obtained in work \cite{10} mainly consists
in the difference between the dependences $\theta_{1}(H_{e})$ and
$\theta_{2}(H_{e})$ for various $\varphi_{\rm H}$.

In Fig.~5, the difference between the populations in both subensembles,
$\Delta p$, in the region of states with inhomogeneous magnetization is shown.
It was calculated, by using Eqs.~(\ref{eq12}) and (\ref{eq14}).

In Figs.\,\,4 and 5, one can clearly see the critical transition
points.\,\,In Fig.~4, these are collapses of two subensembles with
the different directions $\theta_{1}$ and $\theta_{2}$ into a single
ensemble with the direction $\theta m$.\,\,This collapse is the most
pronounced in the case $\theta _{\rm H}=10^{\circ}$; for other
angles, the collapse region narrows and becomes not so
appreciable.\,\,In Fig.~5, the change $\Delta p$ for two
subensembles is rather smooth at small non-zero $\theta _{\rm
H}$-angles.\,\,However, already at $\theta _{\rm H}>10^{\circ}$, the
quantity $\Delta p$ demonstrates a rather strong inflection to the
state with $\Delta p=\pm$ 1 at fields close to the critical one
depending on $\theta _{\rm H}$.\,\,Figure~3 testifies to the
agreement between the experimental values of critical field and
those calculated with the values of $H_{d}^{\max}$, $H_{a\theta}$,
and $H_{a\varphi}$ indicated above.\,\,As follows from Figs.~4 and
5, the values of critical transition field obtained with the same
values of $H_{d}^{\max}$, $H_{a\theta}$, and $H_{a\varphi}$ are in
good numerical agreement with the experimental data shown in Fig.~3.

As to the external field orientation $\theta _{\rm H}=0$ (not
exhibited in Fig.~4), we obtain $\theta_{1}=0$ and
$\theta_{2}=180^{\circ}$ in this case up to $H_{d}^{\max}$.\,\,At
$H=H_{d}^{\max}$, the system transits into a homogeneously
magnetized state, and $\theta_{m}=0.$ In this configuration, the
directions of grain magnetic moments in the subensembles do not
change with the field.\,\,Only the populations of the subensembles
change without any deflection of their moments.\,\,From Fig.~5, it
is evident that, for the orientation $\theta _{\rm H}=0$, the
dependence $\Delta p(H)$ is completely linear at both $\varphi_{\rm
H}$-values.\,\,It saturates at the field equal to the maximum
demagnetizing field $H_{d}^{\max}$.\,\,At other orientations
different from $\theta _{\rm H}=0$, the quantity $\Delta p$ grows
firstly rather slowly in low fields; then, approaching $H_{\rm
crit}$, a drastic increase to $\Delta p=1$ takes place.

As is seen from Figs.\,\,4 and 5, the presence of the in-plane
anisotropy together with  the perpendicular one leads to a relative
shift of the transition points  into the homogeneous state for the
magnetic field lying in the same plane with the easy or hard axis of
the in-plane anisotropy and the film normal.\,\,The value of this
shift depends from $\theta _{\rm H}$. It is absent at $\theta _{\rm
H}=0$, reaches maximum at intermediate values of $\theta _{\rm H}$,
and is equal to $H_{a\theta}-H_{a\varphi}$ at $\theta _{\rm
H}=90^{\circ}$.

If the film has only perpendicular anisotropy (it consists of
uniaxial granules), any in-plane direction is hard.\,\,However, if
the anisotropy has also the in-plane component (if the granules are
biaxi\-al), and if the field $H$ is applied  along the easy axis of
this anisotropy, the transition through the critical point into the
homogeneous state will take place faster than in the former case.

Besides the critical transition field, the in-plane anisotropy also
affects the magnetic susceptibility of granules in low fields,
i.e.\,\,the slope of the curve of the granular contribution to the
dependence $M(H)$ at such fields (see Fig.~2).\,\,From Fig.~6, one
can see how the $\theta _{\rm H}$-dependence of the
susceptibility~-- it is obtained at small fields as the normalized
de\-rivative
\[
\chi_{\rm norm}=\left.  {\frac{\partial\left(  {M/M_{s}}\right)
}{\partial\left(
{H/H_{a\theta}}\right)  }}\right\vert _{H\rightarrow0}%
\]

\noindent -- is modified, by depending on the direction of the
external magnetic field projection on the film plane with respect to
the easy axis of the in-plane anisotropy.\,\,In Fig.~6, the symbols
correspond to experimental data, whereas the curves to the results
of theoretical calculations, with the values of $H_{a\theta}$,
$H_{a\varphi}$ and $H_{d}^{\max}$ being obtained from the results of
approximation of the $H_{\rm crit}$ angular dependences (the
position of the critical transition) given above.\,\,It is possible
to say about quite a good correspondence between the model and the
experiment.

\begin{figure}%
\vskip1mm
\includegraphics[width=7.7cm]{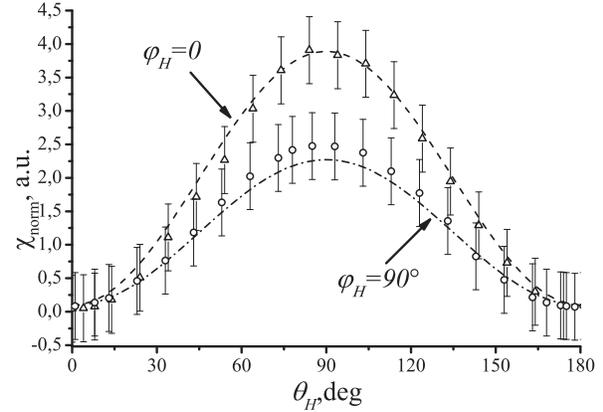}
\vskip-3mm\caption{Comparison between the low-field susceptibilities
measured experimentally (symbols) and the results of model
calculations (curves): $\varphi_{\rm H}=0$ (triangles and dashed
curve) and $\varphi _{\rm H}=90^{\circ}$ (circles and dash-dotted
curve)}\vspace*{-2mm}
\end{figure}

\section{Conclusions}

In works \cite{10,19}, a model for the description of
\textquotedblleft equilibrium\textquotedblright\ dependences
$M(H_{e})$ for NG films with perpendicular anisotropy in a tilted
external magnetic field was considered.\,\,It was verified there for
the ratio $H_{d}^{\max}/H_{a}=3$ and compared with the experiment
for the ratio equal to 4.3.\,\,In this work, this model is compared
with the experiment for the ratios $H_{d}^{\max}/H_{a\theta}\approx$
$\approx24$ and
$H_{d}^{\max}/(H_{a\theta}-H_{a\varphi})\approx41$.\,\,The
comparison results testify that the given model is valid in a wide
interval of values of this ratio.\,\,Naturally, for practical
applications of granular films with perpendicular anisotropy, the
materials, for which this ratio is less than 1, are of high
importance.\,\,The model has no restrictions with respect to the
values of indicated \mbox{ratio.}

The model concerned is developed and compared with the experimental
data obtained for a NG film which, in addition to the perpendicular
anisotropy of granules, also has in-plane anisotropy.\,\,Analytical
expressions in the case of a tilted external field are derived for
the magnetization oriented in planes that contain the normal to the
film and the easy or hard axis of the in-plane anisotropy.\,\,The
problem for an arbitrary magnetization direction can be solved
numerically.

The studied film contained a relative volume of a ferromagnetic
material of 74.5~at.\%, which unambiguously exceeds the percolation
threshold.\,\,However, the magnetization reversal curves show that
the film had attributes typical of a film with perpendicular
anisotropy.\,\,This result was confirmed by the results of
transmission electron microscopy obtained for the same film
\cite{21}.\,\,The peculiarities of the hysteresis sections in the
$M(H_{e})$ curves for the film concerned allowed us to substantiate,
in work \cite{21}, the statement that, in the magnetic aspect, the
examined film is biphase; i.e.\,\,it includes a ferromagnetic
material (Co), some fraction of which forms a large percolation
cluster, and the other fraction consists of electrically insulated
metallic ferromagnetic granules.\,\,A rather large granular fraction
of the film (non-percolating granules) may probably originate from
the fact that Co granules have insulating oxide shells, which
prohibit them from contact with one another.\,\,However, a smaller
fraction of granules form a large percolation cluster.\,\,The model
is developed taking into account that the film has another magnetic
\mbox{component.}

In addition, some specific data for the studied film are
obtained.\,\,The data testifying to its biaxial character are most
interesting.\,\,The biaxial character of the film may probably
result from the used technology of its fabrication under the
conditions of the inclined sputtering on areas with a high content
of the magnetic component.

\vskip3mm \textit{The work was partially supported by the State Fund
for Fundamental Researches of Ukraine in the framework of the
program \textquotedblleft Fundamental problems of nanostructural
systems, nanomaterials, and nanotechnologies\textquotedblright\
(project 1/14-H).}

\rezume{%
М.М.\,Кулик, В.М.\,Калита,\\ А.Ф.\,Лозенко, С.М.\,Рябченко,\\
О.В.\,Стогнєй, А.В.\,Ситніков}{ВПЛИВ ВНУТРІШНЬОПЛОЩИННОЇ\\
АНІЗОТРОПІЇ НА ВЕЛИЧИНУ ПОЛЯ
КРИТИЧНОГО\\
ПЕРЕХОДУ В НАНОГРАНУЛЯРНИХ ПЛІВКАХ\\ З ПЕРПЕНДИКУЛЯРНОЮ
АНІЗОТРОПІЄЮ}{Досліджено вплив внутрішньоплощинної анізотропії на
намагнічування на\-но\-гра\-ну\-ляр\-ної плівки з перпендикулярною
анізотропією. Показано, що в нахиленому до нормалі плівки магнітному
полі спостерігається критичний перехід від неоднорідного магнітного
стану гранул з неколінеарним напрямком їх моментів до однорідного
стану з паралельною орієнтацією магнітних моментів гранул. Отримано,
що внутрішньоплощинна анізотропія впливає на кутову залежність
критичного поля. Теоретичний опис орієнтованого ансамблю двовісних
частинок проведено в наближенні двоямного потенціалу. Незважаючи на
двовісність магнітної анізотропії частинок, у неоднорідному стані,
ансамбль розбивається на два підансамблі, у кожному з яких магнітні
моменти частинок співнаправлені. У критичному полі відбувається
перехід від неоднорідного стану із двома підансамблями до
однорідного стану. Дані розрахунків порівняні з результатами
дослідження наногранулярної плівки Co/Al$_{2}$O$_{n}$, з
перпендикулярною анізотропією, яка містить 74,5 ат.{\%} Co, що
перевищує поріг перколяції. Магнітний момент такої плівки
складається із суми двох внесків: наногранулярного із двовісною
анізотропією гранул і фази, що утворює перколяційний кластер.
Магнітні властивості виділеного із загальної намагніченості
наногранулярного внеску добре узгоджуються з даними розрахунків.}

\end{document}